# Room temperature multiferroism in polycrystalline thin films of gallium ferrite

Monali Mishra[1], Amritendu Roy[2], Ashish Garg[3], Rajeev Gupta[4,5] and Somdutta Mukherjee[1,1]

[1] Colloids and Materials Chemistry Department, CSIR-IMMT Bhubaneswar-751013, India
[2] Minerals, Metallurgical and Materials Engineering, Indian Institute of Technology, Bhubaneswar-751007, India
[3] Materials Science and Engineering, Indian Institute of Technology, Kanpur-208016, India
[4] Department of Physics, Indian Institute of Technology, Kanpur-208016, India
[5] Materials Science Programme, Indian Institute of Technology, Kanpur-208016, India

## Abstract

Sol-gel deposited (010) textured polycrystalline thin films of gallium ferrite ($GaFeO_3$ or GFO) on n-Si(100) and Pt/Si(111) substrates are characterized for room temperature multiferroism. Structural characterization using X-ray diffraction and Raman spectroscopy confirms formation of single phase with nano-sized crystallites. Temperature dependent magnetization study demonstrates ferri to paramagnetic transition at ~300 K. Room temperature piezoresponse force microscopic analysis reveals local 180° phase switching of ferroelectric domains at very high coercive field $E_C$, ~ 1350 kV/cm consistent with recent experimental and first-principles studies. Our study opens up possibility of integrating polycrystalline GFO in novel room temperature multiferroic devices.

Corresponding author: msomdutta@gmail.com

Multiferroics have been recognized as next generation digital memory materials owing to their unique attributes such as lower energy consumption, fast read-write operation and extraordinary memory storage capability. [1, 2] Unfortunately, room temperature multiferroics are rare due to competing requirements of ferroelectricity and magnetism.[3] The most extensively studied room temperature multiferroic, bismuth ferrite ($BiFeO_3$)[4] demonstrates weak/no net magnetization at room temperature and has very weak magnetoelectric coupling limiting its application potential. Thus, search for newer multiferroics with appreciable room temperature properties is a major challenge. Further, commercial device fabrication and design require integration on to a standard silicon circuit such that the material should demonstrate sizable magnetic as well as ferroelectric character along the growth direction. In other words, the material should be grown in such a way that it yields maximum possible multiferroic properties along the growth direction. Thus, an alternative to $BiFeO_3$ for prospective multiferroic memory device applications would be highly desirable. In this context, gallium ferrite is an interesting system with exciting properties.[5]

GFO is a polar piezoelectric with near room temperature ferrimagnetism and magnetoelectric coupling.[6, 7] Further, ferri to paramagnetic transition temperature ($T_C$) can be tailored to room temperature and above by tuning the Ga:Fe ratio slightly away from the stoichiometry.[8] The observed long range magnetic order in the A-type antiferromagnetic ground state [9] could be attributed to cation site disorder between similar sized $Ga^{3+}$ and $Fe^{3+}$ ions wherein some of the $Fe^{3+}$ ions always occupy Ga sites and vice versa, even in stoichiometric composition. [6, 10] Early polarization vs. electric field (P-E) measurement on bulk polycrystalline and single crystalline samples showed leaky dielectric behavior with no/poor signature of ferroelectricity.[11, 12] However, first-principles calculations predicted GFO to undergo ferroelectric to paraelectric phase transition accompanied by a structural transition from polar *Pc2₁n* to centrosymmetric *Pnna* symmetry.[13, 14] Our previous work on epitaxial thin film, in fact, showed that polarization at room temperature could be switched by reversing electric field within nano-dimensional polar domains and thus established GFO as a room temperature ferroelectric[14] and recently confirmed by a study showing saturated ferroelectric P-E loops in epitaxial films of GFO. [15] Ferroelectricity in GFO, coupled with tunable ferrimagnetism at room temperature opens up the possibility of adding a new member to the room temperature multiferroic family, exciting for device applications. However, use of high end deposition technique and expensive single crystalline oxide substrates put a serious question on the commercial efficacy of the material in devices. Therefore, it is imperative to try and deposit GFO on commercial wafer material using low cost deposition techniques. To the best of our knowledge, such an attempt on this system has not been reported so far. In this letter, for the first time, we report observation of nano-scale ferroelectricity as well as room temperature ferrimagnetism in polycrystalline GFO thin films grown using a facile spin coating technique.

Films were spin coated on cleaned n-Si(100) and Pt/TiO$_2$/SiO$_2$/Si(111) or Pt/Si(111) substrates using 0.1 M solution of the precursor nitrates (high purity gallium nitrate hydrate and iron nona-hydrate). Details of synthesis procedure could be found elsewhere.[16] Crystal structure and phase purity of the films were investigated using X-Ray diffraction Cu-Kα radiation. Piezoelectric and ferroelectric behaviors were examined using piezoresponce force microscope (Asylum Research) and Radiant precision premier-II. Temperature dependent and isothermal magnetization were studied with a physical property measurement system (Quantum Design).

Fig. 1(a) shows room temperature XRD patterns of GFO films grown on Pt/Si(111) (GFO||Pt/Si(111)) and n-Si(100) (GFO||n-Si(100)) substrates, respectively, over 2θ range, 18-80˚. XRD patterns were indexed using JCPDS card no. 76-1005 confirming formation of single phase with orthorhombic *Pc2$_1$n* symmetry. One can notice that predominantly (*0k0*) reflections are present. This indicates growth of (010) out-of-plane orientation that coincides the spontaneous polarization direction of GFO.[9, 14] Orthorhombic GFO has lattice parameters *a* = 0.872 nm, *b* = 0.937 nm and *c* = 0.507 nm in bulk structure.[7] For out-of-plane *b*-axis orientation, the in-plane lattice parameters of GFO √(*a$^2$*+*c$^2$*) have a lattice mismatch of ~ 6.8% with twice of lattice parameter of Si. This mismatch is quite large compared to that of GFO grown epitaxially on indium tin oxide (0.7 %) or yttria stabilized zirconia (1.6%).[14] Due to such large mismatch, a large tensile strain is exerted resulting in generation of structural defects which in turn leads to polycrystallinity in the film. As a result, the out-of-plane lattice parameter, *b* ~ 9.34 Å is compressed by only 0.3% compared to its bulk structure as determined from the XRD data. In case of GFO||Pt/Si(111), since the top Pt layer is polycrystalline, heterogeneous nucleation would be more favored at various defect sites such as grain boundaries and our FESEM study in

fact, as discussed below, demonstrates the grain size is significantly less in case of GFO∥Pt/Si(111) compared to GFO∥n-Si(100). We also calculated average crystallite size using Debye-Scherrer's formula: 45 and 36 nm, for GFO∥n-Si(100) and GFO∥Pt/Si(111), respectively.

To investigate the structural behavior, and strain effect, samples were analyzed by Raman spectroscopy using 514 nm $Ar^+$ laser as the excitation source. GFO, with $Pc2_1n$ symmetry has 8 formula units per unit cell resulting in a total of 117 Raman active modes at the zone center.[7] Fig. 1(b) shows unpolarized Raman spectra of GFO films measured in back scattering geometry at room temperature. It is found that Raman spectrum of GFO∥Pt/Si(111) is very similar to that obtained for bulk polycrystalline samples.[8] The peaks are broad in complete agreement with the nanocrystalline nature of the film. We observe a total of 19 modes over 100 to 850 $cm^{-1}$ marked with dotted vertical lines. However, the spectrum of GFO∥n-Si(100) looks different compared to GFO∥Pt/Si(111). Since the penetration depth of 514 nm is larger than the film thickness, intense Raman peak at ~520 $cm^{-1}$ from Si substrate is visible and sample peaks are resolved only after subtracting this intense peak. Though, most of the GFO peaks are present, the spectrum is dominated by other peaks from Si such as at ~ 303 $cm^{-1}$ and 620 $cm^{-1}$.[17,18] Si peaks at 670 $cm^{-1}$ appear as a shoulder to GFO peak at 692 $cm^{-1}$.[17,18] GFO peak at 437 $cm^{-1}$ has been suppressed by a Si peak at 433 $cm^{-1}$.[17,18] Since Raman peak positions of GFO∥Pt/Si(111) and GFO∥n-Si(100) are almost identical, we conclude that the strain levels at both the films are comparable.

FE-SEM micrographs of the GFO thin films, GFO∥Pt/Si(111) and GFO∥n-Si(100) are shown in Fig.2 (a) and (b), respectively. The micrographs reveal a largely uniform deposition of GFO on both of the substrates with larger grains on n-Si(100) compared to on Pt/Si(111). Average grain size of GFO∥Pt/Si(111) and GFO∥n-Si (100) are ~45 nm and 96 nm, respectively. Cross sectional SEM of GFO∥Pt/Si(111) as shown in Fig. 2(c) demonstrates typical film thickness of ~200 nm. The energy dispersive X-ray spectrum analysis (not shown here) revealed that our films are slightly Fe rich with Ga to Fe ratio of 10:12. Surface topography analysis using atomic force microscopy of GFO∥Pt/Si(111) as shown in, Fig. 2(d) reveals a RMS roughness of 1.6 nm bolstering the SEM observation that the films are largely uniform.

Easy magnetization direction of GFO is along crystallographic $c$-axis[6] that lies within our sample plane . Magnetization is studied as a function of temperature in presence of 500 Oe field over 25 K to 320 K. Field cooled (FC) and zero field cooled (ZFC) in-plane magnetization with temperature are plotted in Fig. 3. It was found that GFO undergoes ferri-to-paramagnetic transition at ~ 300 and 302 K for GFO||n-Si(100) and GFO||Pt/Si(111), respectively. These are rather high compared to previously reported $T_C$ of epitaxial film[19] that closely matches with bulk GFO.[6] The higher value of $T_C$ observed could be attributed to the reduced crystallite size of our film,[20] suggesting that the $T_C$ could be adjusted by tailoring the synthesis conditions as well, in addition to sample composition. The maximum magnetization estimated from FC plots ~ 60.1 emu/cc (or 0.33 $\mu_B$/Fe) and ~ 94.6 emu/cc (or 0.53 $\mu_B$/Fe) at 25 K for GFO||Pt/Si(111) and GFO||n-Si(100), respectively are comparable to the value observed in epitaxial thin film.[19] The interesting feature is that the FC and ZFC plots show bifurcation and ZFC magnetization goes negative with decreasing temperature. Negative magnetization is not unusual in ferrimagnetic materials and has also been observed in GFO nanoparticles with crystallite size ranges from 18 to 64 nm.[21] However, for samples with larger grains, the ZFC magnetization is positive over the entire temperature range.[21] GFO contains four types of cation sites: Fe1, Fe2, Ga2 octahedral sites and Ga1 tetrahedral sites with inherent site disorder.[6] In low temperature synthesized GFO,[6] site disorder predominantly affects octahedral sites:[7] $Fe^{3+}$ ions partially occupy octahedral Ga2 sites and Fe sites are partially occupied by $Ga^{3+}$ ions.[6, 9] Fe at Fe1 site couple anti-parallely to Fe at Fe2 and Ga2 sites resulting in the observed ferrinmagnetic order.[6, 7, 9] Magnetization at these different tetrahedral and octahedral sites may have different temperature dependence,[22] Curie temperature ($T_C$), coercive field[21] etc. and the total magnetization is the sum of magnetization at all four sub-lattices.[21, 23] The origin of negative magnetization observed in our films could be attributed to the variation in temperature dependence of magnetization at tetrahedral and octahedral sites coupled to the surface spin structure which changes with synthesis condition. However, in-depth studies are required to understand this effect in GFO nanostructures. Further, magnetization is plotted as a function of magnetic field (M-H) at 299K for GFO||Pt/Si(111) after subtracting the substrate contribution as shown in the inset of Fig. 3. The M-H loop is saturated with saturation magnetization of $M_S$ ~ 5 emu/cc suggesting our polycrystalline film is magnetic at room temperature. The magnitude of magnetization is comparable to those reported for $BiFeO_3$ polycrystalline films at room temperature.[24, 25] However, our previous study[8] and recent report on epitaxial thin films[26] suggest that room temperature magnetization can be further enhanced by tailoring the Fe content in GFO.

For electrical characterization and to explore the potential of GFO in MEMS devices, we chose GFO||Pt/Si(111). The film was sputter coated with Au top electrodes of 0.2 mm diameter to form Pt/GFO/Au capacitors as shown schematically in Fig. 4(a). Fig. 4(b) shows room temperature ferroelectric P-E hysteresis loops at 10 kHz under different electric fields. Form Fig. 4(b) we find that P-E loops do not saturate even at 400 kV/cm and looks more like leaky dielectrics,[27] latter attributed to the conducting grain boundaries masking the intrinsic ferroelectricity of the grains. First-principles calculations predicted that GFO possesses a very high activation energy ~ 0.61 eV/ f. u. to switch from its centrosymmetric to non-centrosymmetric structure which in turn, translates into requirement of a huge coercive field ($E_C$) to switch polarization.[14] In fact, a recent study on GFO epitaxial film reported an $E_C$ of 1400 kV/cm to switch polarization in a *b*-axis oriented film.[15] Due to large leakage in our polycrystalline films, we could not apply such high field.

However, to appreciate the switching behavior of ferroelectric domains, we analyzed the films using piezoresponse force microscopy (PFM). PFM, in addition to visualization of domain structure,[28] allows to induce local polarization switching in individual grains obfuscating the grain boundary contributions whose leakiness otherwise mask the intrinsic ferroelectricity.[29] Fig. 4(c) shows the PFM amplitude image of the as-grown film with domains of different polarity. We chose a square region of 1 μm$^2$ on the sample and sequentially applied -5V, +5V and -5V to study the effect of bias on the domain evolution. Fig. 4(d-f) show the out of plane PFM amplitude images under different applied voltages mentioned above over a scan area of 1.5 μm$^2$. Upon applying -5V, domains with bright shade (apparently with downward polarization) become darker suggesting the change in polarization direction. On the other hand application of +5V transforms darker domains to comparatively brighter, indicating polarization change from upward to downward. Further reversal of voltage brings back the initial polarization state (Fig. 4(f)). Therefore, we find that application of electric field of specific direction can switch the polarization direction in the film. However, it is apparent that application of ±5 V cannot swap the polarization state completely since selected area within the sample did not change from complete dark to complete bright and vice-versa. Above observation only suggests an incomplete polarization switching and indicates requirement of larger $E_C$. We in fact, chose a particular domain and applied higher switching voltages and the effects are shown in switching spectroscopy data plotted in Fig. 5. Fig. 5(a) and (b) plot out-of-plane PFM amplitude and corresponding phase as a function of switching potentials 10 V to 100 V. We find from the phase plot that upon sweeping voltage -10 to +10 V the phase changes about ~ 145° i.e. incomplete switching and with increasing bias voltage phase difference increases. A ~180° phase change occurs at an applied voltage of 100 V as shown in Fig. 5(b) indicating complete polarization reversal. An interesting feature of both Fig. 5(a) and 5(b) is that the loops are not centered at 0 V. They are shifted in the negative voltage axis, instead. If we evaluate the amplitude images (Fig. 4(c-f)) carefully, similar effect could also be observed there. This is the signature of self-biasing in the film and similar effect was previously observed in other ferroelectric thin films as well. [30] Such self-biased behavior could originate from various defects in the film such as oxygen vacancies. These positively charged defects can trap charges and hence generate a local inherent electric field. To compensate this internal field larger amount of electric field is needed to complete the domain switching process. Due to the presence of downward internal field the coercive field becomes -33 V and +21 V on negative and positive side, respectively at complete polarization switching that translates to an average $E_C$ of ~ 1350 kV/cm. Here we note that, this high $E_C$ is in complete agreement to our first-principles study[14] and comparable to the value reported recently on GFO epitaxial thin film determined at macroscopic P-E hysteresis measurement.[15]

In summary, we have, for the first time, synthesized polycrystalline yet oriented GFO thin films with orthorhombic $Pc2_1n$ symmetry using low cost chemical solution deposition technique. The film growth direction coincides with the direction of spontaneous polarization. Though bulk electrical measurement demonstrates an unsaturated P-E hysteresis loop, nano-scale measurement clearly demonstrates switching of polarization states at a high $E_C$ of ~1350 kV/cm, in agreement with recent reports.[15][14] Magnetization measurement reveals a ferri to paramagnetic transition at ~300 K. Saturated room temperature ferrimagnetic loop is observed with weak magnetization. Magnetization value can further be improved by tuning Ga: Fe ratio.[8, 31] Thus, polycrystalline GFO thin films become a potential alternative to other competing systems such as $BiFeO_3$ for designing commercial multiferroic devices. However, the leakage behavior of the films needs to be improved significantly in order to realize room temperature bulk ferroelectric hysteresis loop. In this regard, aliovalent doping at $Fe^{3+}$ site by $Mg^{2+}$ has been proved beneficial as reported in epitaxially grown thin film samples.[31]

This work was supported by Department of Science and Technology, Govt. of India under the INSPIRE-Faculty Award Program through Grant No. IFA13/MS-03. S. Mukherjee and M. Mishra thank Prof. B. K. Mishra, Director of IMMT Bhubaneswar, for providing support for this work. AR thanks IIT Bhubaneswar for research initiation grant through Project No. SP059 and Central Instrumentation Facility at IIT Bhubaneswar for characterization facilities.References:

1. A. Roy, R. Gupta and A. Garg, Advances in Condensed Matter Physics **2012**, 12 (2012).
2. J. F. Scott, Nat Mater **6** (4), 256-257 (2007).
3. N. A. Hill, The Journal of Physical Chemistry B **104** (29), 6694-6709 (2000).
4. G. Catalan and J. F. Scott, Advanced Materials **21** (24), 2463-2485 (2009).
5. A. Roy, S. Mukherjee, R. Gupta, R. Prasad and A. Garg, Ferroelectrics **473** (1), 154-170 (2014).
6. T. Arima, D. Higashiyama, Y. Kaneko, J. P. He, T. Goto, S. Miyasaka, T. Kimura, K. Oikawa, T. Kamiyama, R. Kumai and Y. Tokura, Physical Review B **70** (6), 064426 (2004).
7. S. Mukherjee, A. Garg and R. Gupta, Journal of Physics: Condensed Matter **23** (44), 445403 (2011).
8. S. Mukherjee, V. Ranjan, R. Gupta and A. Garg, Solid State Communications **152** (13), 1181-1185 (2012).
9. A. Roy, S. Mukherjee, R. Gupta, S. Auluck, R. Prasad and A. Garg, Journal of Physics: Condensed Matter **23** (32), 325902 (2011).
10. A. Roy, R. Prasad, S. Auluck and A. Garg, Journal of Applied Physics **111** (4), 043915 (2012).
11. V. B. Naik and R. Mahendiran, Journal of Applied Physics **106** (12), 123910 (2009).
12. Z. H. Sun, S. Dai, Y. L. Zhou, L. Z. Cao and Z. H. Chen, Thin Solid Films **516** (21), 7433-7436 (2008).
13. D. Stoeffler, Journal of Physics: Condensed Matter **24** (18), 185502 (2012).
14. S. Mukherjee, A. Roy, S. Auluck, R. Prasad, R. Gupta and A. Garg, Physical Review Letters **111** (8), 087601 (2013).
15. S. Song, H. M. Jang, N.-S. Lee, J. Y. Son, R. Gupta, A. Garg, J. Ratanapreechachai and J. F. Scott, NPG Asia Mater **8**, e242 (2016).
16. M. Somdutta, Amritendu Roy, in *NMDATM 2016* (IIT Kanpur, India, 2016).
17. A. Hammouda, A. Canizarès, P. Simon, A. Boughalout and M. Kechouane, Vibrational Spectroscopy **62**, 217-221 (2012).
18. J. Kennedy, P. P. Murmu, J. Leveneur, A. Markwitz and J. Futter, Applied Surface Science **367**, 52-58 (2016).
19. K. Sharma, V. Raghavendra Reddy, A. Gupta, R. J. Choudhary, D. M. Phase and V. Ganesan, Applied Physics Letters **102** (21), 212401 (2013).
20. T. C. Han, T. Y. Chen and Y. C. Lee, Applied Physics Letters **103** (23), 232405 (2013).
21. S. Kavita, V. R. Reddy, G. Ajay, A. Banerjee and A. M. Awasthi, Journal of Physics: Condensed Matter **25** (7), 076002 (2013).
22. V. Raghavendra Reddy, K. Sharma, A. Gupta and A. Banerjee, Journal of Magnetism and Magnetic Materials **362**, 97-103 (2014).
23. W. Kim, J. H. We, S. J. Kim and C. S. Kim, Journal of Applied Physics **101** (9), 09M515 (2007).
24. Y.-H. Lee, J.-M. Wu and C.-H. Lai, Applied Physics Letters **88** (4), 042903 (2006).
25. Z. Quan, W. Liu, H. Hu, S. Xu, B. Sebo, G. Fang, M. Li and X. Zhao, Journal of Applied Physics **104** (8), 084106 (2008).

26. A. Thomasson, S. Cherifi, C. Lefevre, F. Roulland, B. Gautier, D. Albertini, C. Meny and N. Viart, Journal of Applied Physics **113** (21), 214101 (2013).
27. J. F. Scott, Journal of Physics: Condensed Matter **20** (2), 021001 (2008).
28. H. Yoo, C. Bae, M. Kim, S. Hong, K. No, Y. Kim and H. Shin, Applied Physics Letters **103** (2), 022902 (2013).
29. S. V. Kalinin and D. A. Bonnell, Physical Review B **65** (12), 125408 (2002).
30. P. Miao, Y. Zhao, N. Luo, D. Zhao, A. Chen, Z. Sun, M. Guo, M. Zhu, H. Zhang and Q. Li, Scientific Reports **6**, 19965 (2016).
31. C. Lefevre, R. H. Shin, J. H. Lee, S. H. Oh, F. Roulland, A. Thomasson, E. Autissier, C. Meny, W. Jo and N. Viart, Applied Physics Letters **100** (26), 262904 (2012).

Figure captions:

Fig.1 (Color online) (a) (a) Indexed X-ray diffraction patterns of (0$k$0) oriented GFO films grown on Pt/Si (111) and n-Si (100) substrates. Presence of a weak GFO (221) peak is marked. A few unidentified peaks originating from the substrates are marked with asterisks. (b) Raman spectra of GFO films grown on Pt/Si (111) and n-Si (100) substrates. Raman scattering peaks originating from Si substrate are marked with arrows in Raman spectrum of GFO on n-Si substrate.

Fig.2 (Color online) FE-SEM images of GFO films deposited on (a) Pt/Si (111) and (b) n-Si (100) substrates. (c) Cross-sectional SEM image of GFO films deposited on Pt/Si (111), (d) 3-D AFM image of GFO film deposited on Pt/Si (111)

Fig. 3 (Color online) FC and ZFC magnetization at 500 Oe as a function of temperature on GFO films deposited on Pt/Si(111) and n-Si(100) substrates. Inset shows room temperature magnetization vs. magnetic field for GFO grown on Pt/Si(111) substrate.

Fig. 4(a) (Color online) Schematic of MIM capacitor geometry used for ferroelectric measurement, (b) Room temperature ferroelectric hysteresis loop of GFO film deposited on Pt/Si (111) substrate measured at 10 kHz; (c)-(f) Out of plane PFM images of Au/GFO/Pt capacitors at applied ±5 V.

Fig. 5 (Color online) Piezoelectric switching spectroscopy measurement on a particular piezoelectric domain: piezoelectric amplitude and phase hysteresis loops upon reversal of several bias voltages.

Figures

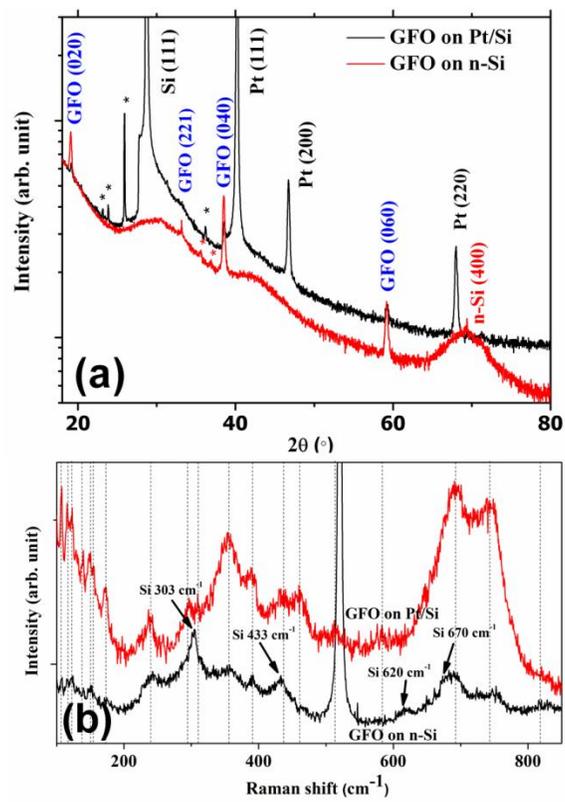

Fig. 1

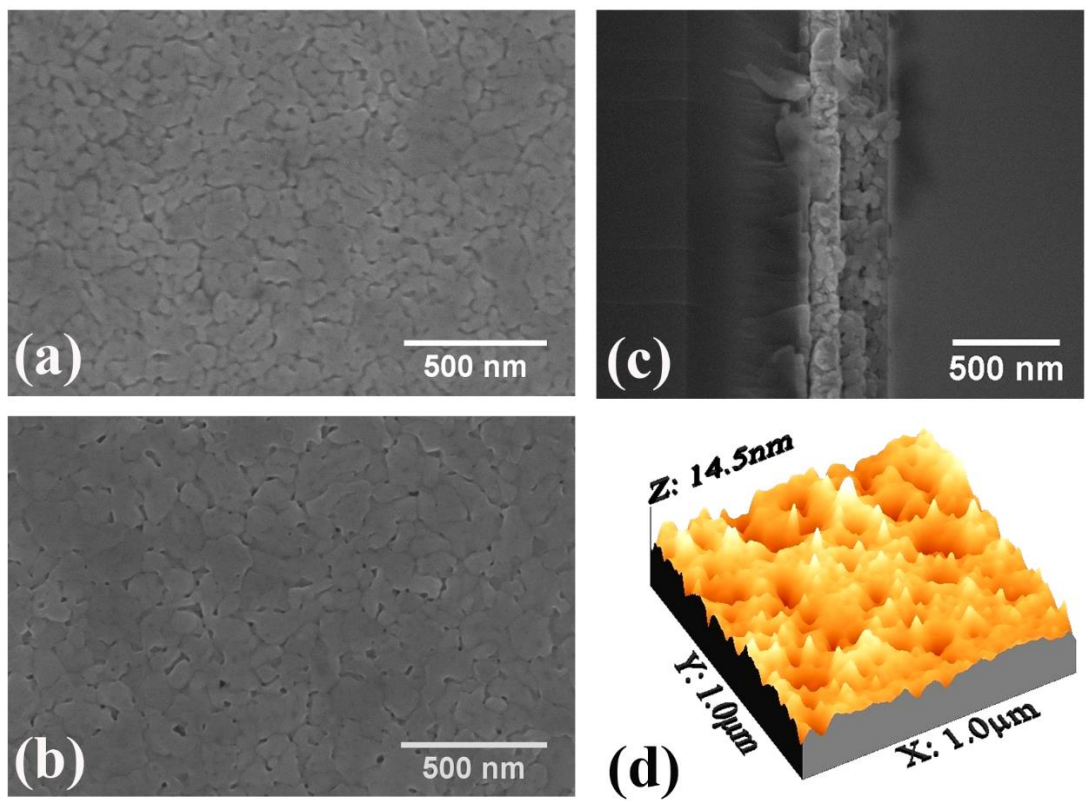

Fig. 2

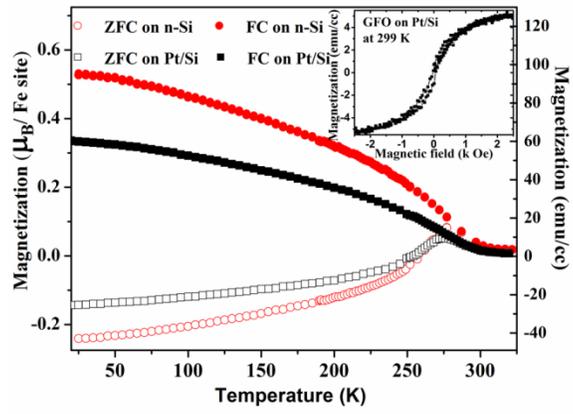

Fig. 3

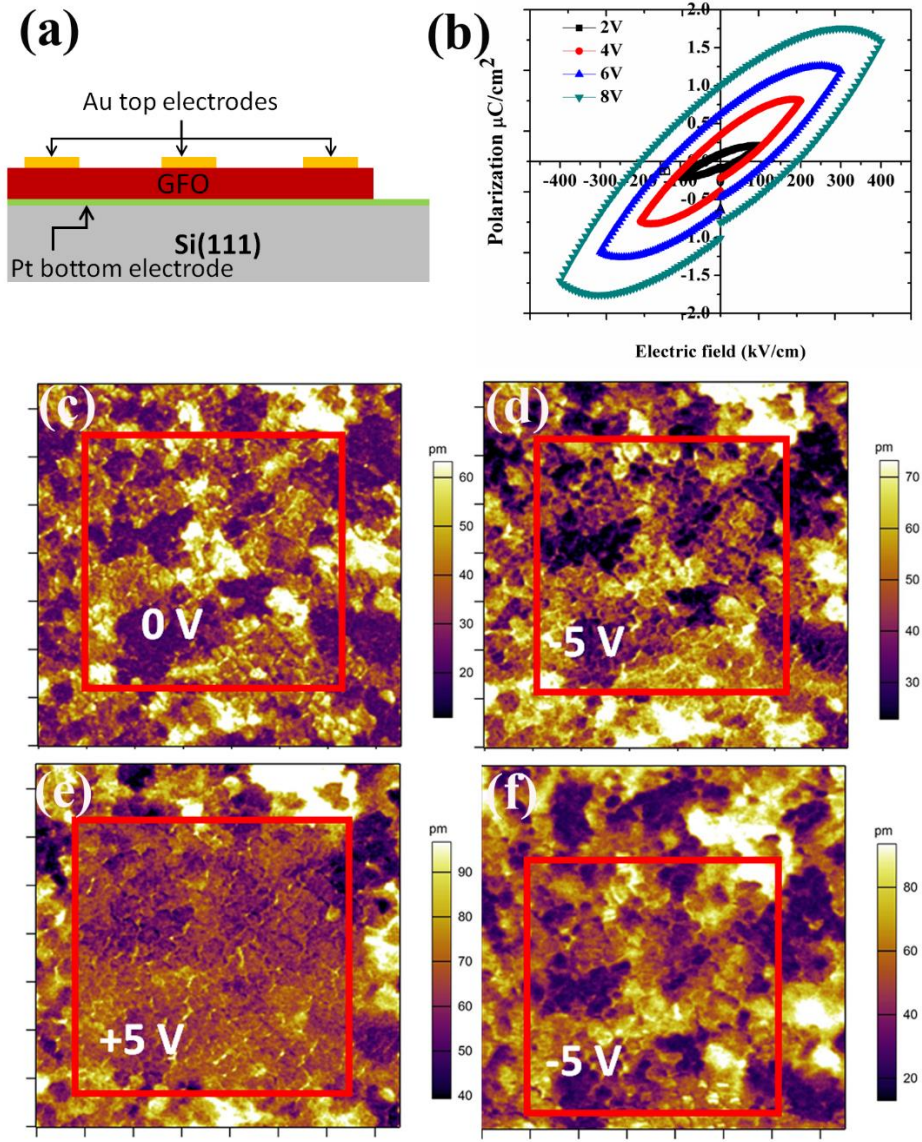

Fig. 4

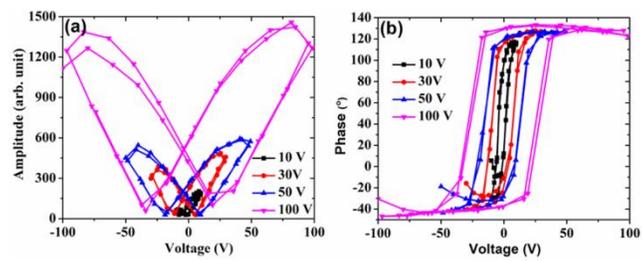

Fig. 5

1515